\documentclass[runningheads]{llncs}
\usepackage[T1]{fontenc}
\usepackage{graphicx}
\begin{document}

\title{Optimizing Predictive Maintenance: Enhanced AI and Backend Integration}
% If the paper (1)title is too long for the running head, you can set

\author{Michael Stern\inst{1}\orcidID{0009-0002-7558-5384}
\texttt{michael.stern@hshl.de}\and
Michelle Hallmann\inst{1}\orcidID{0009-0002-9600-7575} \and
Francesco Vona\inst{1}\orcidID{0000-0003-4558-4989}\and
Ute Franke\inst{2}\orcidID{0009-0009-3945-4775} \and
Thomas Ostertag\inst{3}\orcidID{0009-0005-8617-3256 } \and
Benjamin Schlüter\inst{4}\orcidID{0009-0005-0374-363X} \and
Jan-Niklas Voigt-Antons\inst{1}\orcidID{0000-0002-2786-9262}
}

\authorrunning{Stern et al.}

\institute{Hamm-Lippstadt University of Applied Sciences,  59557 Lippstadt, Germany \and
5micron GmbH, 12489 Berlin, Germany \and
OSTAKON GmbH, 10999 Berlin, Germany \and
Deutsche Eisenbahn Service AG, 16949 Putlitz, Germany}
\maketitle              % typeset the header of the contribution

\noindent
This version of the contribution has been accepted for publication, after peer review. The Version of Record is available online at: https://doi.org/10.1007/978-3-031-78531-3\_30

\begin{abstract}

Rail transportation success depends on efficient maintenance to avoid delays and malfunctions, particularly in rural areas with limited resources. We propose a cost-effective wireless monitoring system that integrates sensors and machine learning to address these challenges. We developed a secure data management system, equipping train cars and rail sections with sensors to collect structural and environmental data. This data supports Predictive Maintenance by identifying potential issues before they lead to failures. Implementing this system requires a robust backend infrastructure for secure data transfer, storage, and analysis. Designed collaboratively with stakeholders, including the railroad company and project partners, our system is tailored to meet specific requirements while ensuring data integrity and security. This article discusses the reasoning behind our design choices, including the selection of sensors, data handling protocols, and Machine Learning models. We propose a system architecture for implementing the solution, covering aspects such as network topology and data processing workflows. Our approach aims to enhance the reliability and efficiency of rail transportation through advanced technological integration.

\end{abstract}

\section{Introduction}
The economic success of rail transportation significantly hinges on the costs associated with maintaining infrastructure and vehicles \cite{ref_Maintenance}. Inadequate maintenance can lead to train delays and malfunctions \cite{ref_TrainDelays}, resulting in increased public dissatisfaction with the public transportation system \cite{ref_ValuationReliabiltyTravel}\cite{ref_UnreliablityPublicTransport}. Therefore, finding cost-efficient solutions to enhance transport capacity in rural areas and the overall appeal of rail transportation is essential. One promising approach involves deploying affordable wireless monitoring systems for rail infrastructure and vehicles, utilizing cost-effective sensors and Machine Learning (ML) algorithms \cite{ref_DegradationPrediction}.
These systems must be designed with low-threshold retrofit solutions, allowing seamless integration into existing infrastructure. Additionally, they should adopt a systemic approach that considers the entirety of rail interactions, emphasizing the importance of data integrity and security. To effectively integrate Predictive Maintenance (PdM) systems into rail transportation maintenance, collecting accurate data for informed maintenance decisions is crucial, as the capabilities of a Machine Learning system are heavily dependent on the quality of the training data \cite{ref_DataQualityML}, while also implementing robust security measures to safeguard sensitive rail infrastructure information. Hence, developing a backend infrastructure that is adaptable and scalable for this purpose is imperative. While guidelines for backend infrastructures, including databases and AI, are essential, they must be tailored to meet the specific requirements of this field for successful integration. 
This article presents the design of a backend infrastructure for integrating a monitoring and Predictive Maintenance system for trains and rail infrastructure in a rural area of Germany. The project aims to evaluate the effectiveness of structure-borne noise measurement methods for various monitoring and maintenance tasks on rails and vehicles. 
For this purpose, two train cars and one rail section are equipped with structure-borne noise measurement sensor systems developed by the project partner responsible for the sensory hardware. These sensors are general-purpose, cost-effective, and highly scalable. In addition to recording structure-borne noise, the sensors also capture other data such as GPS coordinates, temperature, timestamps, and acceleration in the XYZ axis, as well as roll, pitch, and yaw. By leveraging the unique capability of acoustics to derive specific findings from non-specific signals, the system detects the condition and material properties of vehicles and infrastructure. The sensors in the train cars monitor the state of the rails, while the sensor on the rail section monitors the condition of the train cars as they enter and leave the railway workshop. During train rides, events are labeled by the train drivers, and the workshop mechanics label faults and repairs to the vehicles. 
A distributed ledger technology (DLT) elevates the infrastructure to the web3 standard, ensuring essential use cases are accessible across companies. This DLT network facilitates decentralized data exchange and transfers labeling data to a data analysis server. Here, Machine Learning algorithms process the labels and training data to generate recommended measures for the workshop foreman. This technology provides numerous advantages in terms of security and data integrity. Once data is collected and written into the network, it becomes immutable and resistant to manipulation. Any work or analysis conducted on the data and adding fields to the dataset is preserved as a new version rather than replacing the original. Such an approach is particularly crucial in safety-critical systems like railways, where precise and dependable data is vital for operations, maintenance, and compliance with regulatory standards.
A distributed ledger network enhances transparency and traceability by preserving the original data and maintaining a comprehensive history of all changes. This immutability and version control ensure that all recorded data remains tamper-proof, providing a robust foundation for trust and accountability in the rail-critical environment.
Furthermore, this network transfers the recorded and collected data between different servers equipped with storage, analysis, and visualization tools. By presenting the processes and design implementation of our project, we aim to provide suggestions for designing backend infrastructures in the area of Industry 4.0, with a specific focus on integrating Predictive Maintenance systems in the rail transportation industry.

\section{Related Work}
\subsection{Predictive Maintenance}
Recent works have increasingly utilized Machine Learning and Deep Learning (DL) for Predictive Maintenance in railways, focusing on analyzing sensor data to detect anomalies and predict faults. The goal is to optimize maintenance activities and reduce costs, with various surveys highlighting the effectiveness, challenges, and future research directions of these ML and DL methods \cite{ref_DataDriven}. Among the ML models, tree-based classification techniques like decision trees and random forests have shown particular promise. These models can predict maintenance needs using existing data from railway agencies, offering understandable results and adaptability to various asset types and planning scenarios \cite{ref_Tree}.

Predictive models built from extensive historical data, including detector and failure data, also play a crucial role in preventing service interruptions and improving network velocity. These models use correlation, causal, time series analysis, and ML techniques to predict failures and optimize maintenance schedules \cite{ref_Velo}. Additionally, emerging trends and techniques in Predictive Maintenance enhance reliability, safety, availability, efficiency, and quality across various sectors while reducing environmental impact \cite{ref_Trends}. Advanced technologies like the Internet of Things (IoT) and radio-frequency identification (RFID) have facilitated more efficient PdM applications, making them increasingly prevalent in industries \cite{ref_Trends}.

\subsection{Condition Monitoring}
In the railway industry, transitioning from time or distance-based maintenance to Predictive Maintenance is essential for reliable fault detection \cite{ref_Condition}. This transition addresses challenges in identifying faults promptly and satisfying maintenance needs cost-effectively. Active condition monitoring is crucial to achieve this, necessitating a shift from visual inspections and preventive measures to Predictive Maintenance strategies. The exploration delves into condition monitoring, particularly rail vehicle suspension and track conditions. Using cost-effective sensors, efficient monitoring methods are proposed and evaluated \cite{ref_Condition}. Experimental and simulation results demonstrate promising approaches, including cross-correlation for suspension fault detection \cite{ref_Condition}. Additionally, vibration-based condition monitoring plays a significant role in civil engineering, offering non-destructive sensing and analysis to detect changes indicating damage or degradation in infrastructure. By reviewing state-of-the-art vibration-based monitoring, the potential for more economical infrastructure management and maintenance is highlighted \cite{ref_Vib1}. Furthermore, vibration monitoring is preferred for fault diagnosis in rotating machinery due to its ability to detect dynamic components' manifestations in measured vibration responses. The basics of vibration monitoring for generic machines are outlined, emphasizing fault detection through signal analysis \cite{ref_Vib2}. Lastly, excessive noise often signifies defects in operating machinery. Signal analysis reveals the defect and its cause, making it crucial for effective monitoring. This discussion of noise basics and its relation to machine defects is essential for comprehensive condition monitoring strategies \cite{ref_Vib2}.

\subsection{Cybersecurity in Railway Digitalization}
Digitalization is transforming the railway industry globally, bringing significant advantages and increased exposure to cyberattacks \cite{ref_Cyber1}. As the number of digital items and interfaces between digital and physical components in railway systems grows, new frameworks, concepts, and architectures are needed to ensure resilience against cybersecurity challenges. These challenges include the lack of proactiveness, a holistic perspective, and the obsolescence of safety systems exposed to current and future cyber threats \cite{ref_Cyber1}. Several studies have examined cybersecurity aspects and their application to railway infrastructure, but a comprehensive roadmap is required to address emerging challenges and solutions effectively \cite{ref_Cyber1}.
The swift advancement of digital technology within railway infrastructure has sparked a research trend focused on data-driven decision-making algorithms. These algorithms offer numerous advantages for the efficient and effective design, construction, operation, and maintenance of railway systems \cite{ref_Cyber2}. However, the breach of sensitive data poses significant risks, including trust, hazards, and severe consequences for railway infrastructure. This underscores the need for robust data security measures within electronic maintenance, emphasizing the importance of managing potential challenges, threats, and vulnerabilities \cite{ref_Cyber2}. Therefore, ensuring data security is critical as railways transition into the digital era, highlighting the need for secure data-driven approaches to maintain continuous operation, efficient maintenance, and informed planning and investments \cite{ref_Cyber2}.
Technological advances in telecommunications have particularly benefited the railway industry, enhancing the management and performance of communication networks. However, these interconnected systems are highly susceptible to cyberattacks \cite{ref_Cyber3}. Cybersecurity standards, guidelines, frameworks, and technologies are used to assess and mitigate risks, especially concerning the relationship between safety and security \cite{ref_Cyber3}. Special attention is given to signaling systems, which rely heavily on computer and communication technologies \cite{ref_Cyber3}. The use of cyber ranges to model and emulate computer networks and attack-defense scenarios is also discussed, along with several use cases relevant to the railway industry \cite{ref_Cyber3}. Integrating advanced technologies necessitates a comprehensive understanding of the security measures required to protect these systems effectively \cite{ref_Cyber3}.
The increasing frequency and severity of cyberattacks across various sectors, including railways, pose significant threats to the safety of employees, passengers, and the public. These attacks result in the loss of sensitive information, reputational damage, and financial losses. Advanced security analytics and automation are crucial for identifying, responding to, and preventing security breaches \cite{ref_Cyber4}. To address these challenges, the authors of \cite{ref_Cyber4} proposed an adapted version of the Cybersecurity Capability Model (C2M2) for railway organizations called Railway-Cybersecurity Capability Maturity Model (R-C2M2). This model includes and integrates advanced security analytics and threat intelligence. The R-C2M2 was used to evaluate the cybersecurity maturity of three railway organizations, providing them with recommendations and action plans for improvement \cite{ref_Cyber4}.

\section{System Architecture}
\subsection{Understanding the current system for maintenance}
Before designing and implementing a Predictive Maintenance system tailored to the railroad company's requirements, delving into the current system and comprehending the existing steps and processes was imperative. This involved conducting interviews with experts and personnel responsible for registering potential malfunctions of vehicles and tracks and gaining insights into current processes and maintenance procedures. Additionally, visits to workshops and discussions with mechanics and officials from the railroad company provided a comprehensive understanding of how maintenance is currently handled, laying the groundwork for developing an appropriate system for transferring and analyzing the data. Methods for relaying newly obtained information back to relevant positions within the maintenance chain were also devised.

\subsection{Conception of a suitable system for Predictive Maintenance}
To develop an effective Predictive Maintenance system, we advocate for a collaborative approach involving the expertise of all project partners. The solution will utilize sensor systems developed by one of the project partners to capture structure-borne noise. There will be two types of sensors deployed:
\begin{itemize}
    \item The first type of sensor is designed to be installed on the vehicle's underbelly on the accelerating axle to record data from the tracks, axles, and engine without potential interference from the wagon chassis. These sensors will record structure-borne noise and capture acceleration on the XYZ axis, pitch, roll, and yaw. Additionally, they will gather GPS data, accurate time stamps, and temperature readings during normal operations. This comprehensive data collection will enable the system to detect anomalies on the tracks, such as bumps, flat spots, and other common issues.
    \item The second type of sensor will be embedded within the tracks, positioned strategically in front of workshops. This setup will ensure that train vehicles pass over the sensors twice—once before maintenance and once after—providing a clear comparison of sound signatures pre- and post-maintenance without external interference. Each sensor unit will consist of two subunits placed on each side of the track and spaced apart to discern the direction of the train's movement by comparing the signals. Observing the time difference between the signals will determine the direction of travel. Additionally, these sensors will record temperature and accurate time stamps.
\end{itemize}

Collecting as much data as possible to train a Machine Learning algorithm will be crucial in the development phase.
This data necessitates clear labeling for distinguishing between events and malfunctions. To enable this, a system will be deployed to capture voice messages from onboard personnel, especially the train driver, whenever unusual incidents occur during routine operations. These voice recordings act as reminders, assisting drivers in recalling daily occurrences and refining the accuracy of collected information for labeling purposes. Notably, these recordings will specifically document data collected by the vehicle. Furthermore, drivers review the recordings after their shift to complete event forms for labeling. This post-shift process is designed to avoid disrupting daily operations and is also implemented for security reasons, as drivers are prohibited from using phones or computers while driving.
For the data collected by the sensors in front of the workshops, mechanics, particularly the head mechanic, will be instructed to label malfunctions and defects upon the vehicle's entry and document the vehicle's state upon exit. This will ensure that the dataset reflects the work performed on damaged vehicles. Vehicles entering and leaving the workshop for checkups will also undergo this process, maximizing the data collected.

\subsection {Data and Algorithms}
The collected data undergoes several key processes. Initially, the raw data is stored on a hard drive, which is retrieved weekly by personnel from the sensor development company. This data is then transferred to the company's servers for further analysis and troubleshooting. Concurrently, the raw data undergoes compression via a computational unit within the sensors. This compression involves preprocessing the sound signal from each reading using Fast Fourier Transformation (FFT) \cite{ref_fft}. This transformation is vital to reduce dataset sizes, allowing smooth transmission via 4G mobile networks. The preprocessed data is then sent to storage servers using a ledger client provided by the project's decentralized ledger technology partner.
Upon arrival at the server cluster, the data is integrated into the blockchain, ensuring immutability.
The analysis servers retrieve the recorded preprocessed data from the cluster for further processing. This process is facilitated by a dockerized container infrastructure that manages and scales processing tasks efficiently. Within this infrastructure, a Flask container hosts a Python application responsible for managing traffic and handling requests, enabling seamless communication between system components. Meanwhile, an Nginx container acts as a web server and reverse proxy server, deploying the labeling interface to train employees. This intuitive platform allows for labeling events and documentation of vehicle states before and after maintenance. A PostgreSQL Database container securely stores sorted data, including labels, events, and metadata, forming an interconnected and resilient processing pipeline that ensures system operation and efficacy while maintaining security through credential management.
\begin{figure}[ht]
    \centering
    \includegraphics[width=1\linewidth]{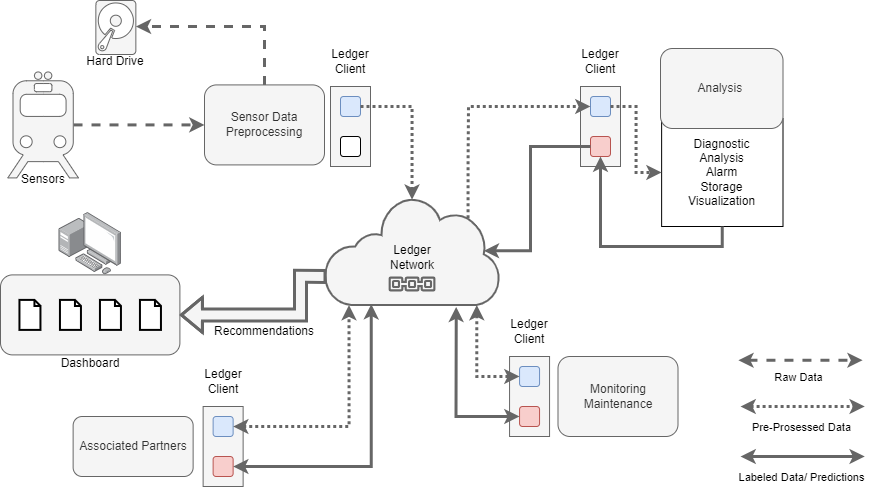}
    \caption{
The figure illustrates the integrated system for vehicle and track diagnostics, monitoring, and maintenance. Sensors on the vehicle and tracks collect raw data stored on a hard drive. The preprocessed data is sent to a decentralized ledger network via a ledger client. This network facilitates secure data sharing and analysis, covering diagnostics, alarm triggering, storage, and visualization. The analysis results and labeled data are used to maintain the vehicle and tracks, with recommendations displayed on a dashboard for associated partners and stakeholders.}
    \label{labelfig}
\end{figure}

As the data is sorted into databases, additional labels, events, and metadata fields are added to facilitate efficient querying and retrieval. An online labeling system is implemented, allowing on-site staff, including train drivers and workshop mechanics, to provide detailed annotations and context crucial for training Machine Learning algorithms. These labels are pivotal in identifying patterns and predicting potential issues. To ensure actionable insights are readily available, robust data visualization and reporting tools generate reports and dashboards summarizing the system's health, highlighting areas requiring attention. Workshop mechanics and stakeholders leverage these visualizations to make informed maintenance schedules and intervention decisions.
After the analysis server processes the data, it is stored in the ledger network. From there, associated partners and responsible railroad company staff members can access the data for further analysis or interpretation.
Once the Machine Learning algorithm is implemented and trained with the collected data, the results will be pushed again into the ledger network. Designated endpoints will then receive recommendations for repairs or alerts on potential malfunctions.

Summarizing, the ledger network will encompass three main types of data:
\begin{itemize}
    \item The original preprocessed data: fast Fourier transformed raw data recorded and collected by the sensor-systems
    \item The labeled and enriched data: FFT Data enriched with various additional information provided by responsible staff members
    \item The final predictions and recommendations: Data analyzed through ML with a prediction on possible malfunctions and repairs
\end{itemize}
The complete system architecture is presented in Figure \ref{labelfig}

\subsection{Security}
Security is a paramount consideration in the design of the backend infrastructure, particularly given the railway-critical environment and the importance of the data it handles. A comprehensive security framework will be implemented to safeguard sensitive information and ensure compliance with industry standards.
Encryption protocols will protect data at rest and in transit, mitigating the risk of unauthorized access or interception. Access controls will restrict system access to authorized personnel only, reducing the potential for data breaches or tampering. Additionally, robust audit trails will be established to track system activity, providing visibility into unauthorized access attempts or anomalous behavior.

\section{Conclusion}
This article presents the design and implementation of a backend infrastructure suitable for integrating machine learning into monitoring and predictive maintenance systems for railroad infrastructures. The primary aim of this project is to develop a system that enhances rail transportation's economic and operational viability. By recording structure-borne noise and environmental data and leveraging this information for predictions through machine learning, our goal is to achieve reliable and precise predictions of potential malfunctions and errors.
The design of this backend infrastructure was heavily influenced by the need for security and data immutability, which are crucial for the rail-critical environment. Upholding these security standards is essential, as the information and results generated are critical for safety and efficiency.
Our system is prepared to receive the preprocessed data. Sensors are currently under deployment, and once the data starts to be collected and labeled, the machine learning algorithms can be trained.
In conclusion, scalability and cost-effectiveness are core objectives in the ongoing development of the backend infrastructure for railway monitoring and predictive maintenance described in this article. If initial testing proves successful and predictions demonstrate high precision, we intend to scale the system to larger use cases while ensuring affordability. Future efforts will focus on validating maintenance predictions with the assistance of responsible companies and verifying the accuracy of train malfunction predictions through thorough checks by mechanics. Through collaborative efforts and rigorous validation processes, we aim to create a scalable, cost-effective, and robust predictive maintenance system that enhances railway operations' reliability, safety, and efficiency. 

\subsubsection{Acknowledgments.} We gratefully acknowledge financial support through the TÜV Rheinland Consulting GmbH with funds provided by the Federal Ministry for Digital and Transport (BMDV) under Grant No. 19F2265 (DigiOnTrack). We also want to thank the reviewers for their thoughtful feedback.

In this paper, we used Overleaf’s built-in spell checker, the current version of ChatGPT (GPT 3.5), and Grammarly. These tools helped us fix spelling mistakes and get suggestions to improve our writing. If not noted otherwise in a specific section, these tools were not used in other forms.

\subsubsection{Disclosure of Interests.}
The authors have no competing interests to declare relevant to this article's content. 

 \bibliographystyle{splncs04}
 \bibliography{main}

\end{document}